\documentclass[twocolumn]{aastex631}

\newcommand{\sext}{{\tt SExtractor}}

\begin{document}

\title{Potential Nitrogen Enrichment via Direct-Collapse Wolf-Rayet Stars in a $z=4.7$ Star-Forming Galaxy}

\correspondingauthor{Yechi Zhang}
\email{yczhang@icrr.u-tokyo.ac.jp}

\author[0000-0003-3817-8739]{Yechi Zhang}
\affiliation{IPAC, California Institute of Technology, MC 314-6, 1200 E. California Boulevard, Pasadena, CA 91125, USA}

\author[0000-0002-8512-1404]{Takahiro Morishita} 
\affiliation{IPAC, California Institute of Technology, MC 314-6, 1200 E. California Boulevard, Pasadena, CA 91125, USA}

\author[0000-0001-9935-6047]{Massimo Stiavelli}
\affiliation{Space Telescope Science Institute, 3700 San Martin Drive, Baltimore, MD 21218, USA}

\begin{abstract}
We present analyses of a nitrogen-enriched star-forming galaxy, ID60001, at $z=4.6928$ based on JWST/NIRSpec MSA spectroscopy and NIRCam photometry. From rest-frame optical emission lines we derive the nitrogen-to-oxygen (N/O) abundance ratio of ID60001 to be $\log({\rm N/O})=-0.76_{-0.03}^{+0.03}$ ($[{\rm N/O}]=0.10_{-0.03}^{+0.03}$), which is significantly elevated at the corresponding metallicity $12+\log({\rm O/H})=7.75_{-0.01}^{+0.01}$ ($Z/Z_\odot = 0.12$) compared to local counterparts. We discuss possible scenarios for elevated N/O abundance in ID60001, including pristine gas inflow, Wolf-Rayet (WR) stars, and Oxygen depletion by Type II supernova winds. Based on the moderately broadened He{\sc ii}$\lambda$4686 emission line, galaxy morphology, and star-formation history, we conclude that the elevated N/O abundance of ID60001 is likely originated from massive ($>25\,M_\odot$) WR stars that directly collapse into a black hole. We also stress the importance of reliable electron density measurements when deriving N/O abundance with rest-frame optical emission lines.
\end{abstract}


\section{Introduction} \label{sec:intro}
Chemical abundance ratios of the interstellar medium (ISM) in early galaxies are keys to understanding galaxy formation processes. Stars with different masses and lifetime undergo different nucleosynthesis processes, resulting in enrichment of various elements. As such, abundance ratios can reveal detailed star formation processes in early galaxies. 

With the James Webb Space Telescope (JWST), it is now possible to directly probe the ISM of $z>4$ galaxies. Previous results from chemical abundance ratio studies with JWST have shown surprising departures from local relations. For example, the N/O abundance ratio of star-forming galaxies in the local Universe increases with gas-phase metallicity \citep[e.g.,][]{pilyugin12,kojima17,berg19,berg20}, which is likely explained by the primary Oxygen production by massive stars and a secondary Nitrogen enrichment from low- and intermediate-mass stars that evolve into the asymptotic giant branch (AGB) phase. In contrast, an elevated N/O abundance at relatively low gas-phase metallicity has been observed at high-$z$ in GN-z11\citep[e.g.,][]{bunker23,cameron23} and other galaxies \citep[e.g.,][]{isobe23c,mc24,maiolino24,topping24} based on rest-frame ultraviolet (UV) lines. Possible origins of such an elevated N/O ratio remain under debate, sometimes requiring exotic sources such as supermassive stars \citep[SMS;][]{mc24},  or fine-tuned intermittent star formation histories involving Wolf-Rayet (WR) stars \citep[e.g.,][]{kobayashi24}.

Another unsolved issue regarding the elevated N/O abundance at high-$z$ is that the N/O abundance of these high-$z$ samples are estimated from high-ionization UV emission lines instead of the low-ionization optical emission lines used in the local galaxies \citep[e.g.][]{pilyugin12,berg19,berg20}. The UV and optical lines may trace different regions of the ISM, hence resulting in inconsistent N/O measurements even for the same objects \citep{ji24}. \citet{stiavelli24} performed a direct comparison of optically-derived N/O abundance between $z>4$ and local galaxies based on data collected from JWST GTO-1199 and 2758 programs and galaxies compiled from the literature \citep{sanders24,rogers24}, finding that three out of 12 galaxies in their sample observed elevated N/O abundance. However, the possible origins of such a high N/O abundance in these galaxies remain unclear because of the restricted ISM information available from the limited wavelength coverage and signal-to-noise (S/N) of NIRSpec MSA data.

In this paper, we aim to distinguish different possible scenarios of N/O enrichment at high-$z$ by revisiting one of the \citet{stiavelli24} sample galaxies, ID60001 at $z=4.693$, which was originally not classified as a N/O enhanced object but has potential outflow features in the NIRSPec MSA. Thanks to the exceptionally good S/N of NIRSpec MSA data, we perform a comprehensive ISM analysis, deriving the updated ISM properties and chemical abundances that yield to a high $\log({\rm N/O})=-0.76_{-0.03}^{+0.03}$ at $12+\log({\rm O/H})=7.75_{-0.01}^{+0.01}$. Complementing with photometry analysis based on NIRCam imaging data, we discuss the possible origin of high N/O in ID60001 based on its ISM and galaxy properties.
%Past studies: \citet{isobe23c}; \citet{schaerer24}; GN-z11 \citep{bunker23,cameron23}, \citet{mc24}: WR wind or SMS; \citet{topping24}: high density, bursty SF; \citet{ji24}: different N/O from UV and optical lines, indicating dense region has N-enrichment first.
This paper is organized as follows. Section \ref{sec:data} summarizes the JWST NIRCam photometry and NIRSpec Multi-Object Spectroscopy (MSA) data used in this study. We present the ISM and host galaxy analysis based on NIRSpec MSA and NIRCam photometry in Section \ref{sec:ism} and \ref{sec:host}, respectively. In Section \ref{sec:discuss}, we discuss the possible origins of the enhanced Nitrogen abundance in ID60001 and the potential effect of density measurements on N/O abundance estimation. Our main findings are summarized in Section \ref{sec:summary}.

\section{Observations and Data} \label{sec:data}

\subsection{NIRCam photometry}\label{subsec:photom}

We utilize reduced NIRCam imaging taken in the 1199-par field, i.e. NIRCam-parallel field taken as part of the GTO program 1199 in JWST Cycle 1 \citep{stiavelli23}. The data were reduced in \citet{morishita24sm}; briefly, we started with the raw-level images from the Mikulski Archive for Space Telescopes (MAST) archive and then reduces them with the official JWST pipeline. Our reduction adds several custom steps including $1/f$-noise subtraction using {\tt bbpn}\footnote{\url{https://github.com/mtakahiro/bbpn}}, snowball masking using {\tt Grizli} \citep{brammer22}, and additional cosmic-ray masking using {\tt lacosmic} \citep{vandokkum01,bradley23}. The final drizzled images, with the pixel scale set to $0.\!''0315$\,/\,pixel, are aligned to the IR-detection image (F277W+F356W+F444W as the default choice). Filters and limiting magnitudes ($5\,\sigma$) are: NIRCAM-F410M (28.8), NIRCAM-F444W (28.8), NIRCAM-F150W (29.2), NIRCAM-F356W (29.3), NIRCAM-F200W (29.4), NIRCAM-F277W (29.3), NIRCAM-F115W (29.1), NIRCAM-F090W (29.2).

Photometry is performed in the consistent manner as \citep{morishita24a}. We identify sources in the detection image using \sext\ \citep{sex96}. We measure source fluxes using PSF-matched (to F444W), with a fixed aperture of radius $0.\!''16$. We set the configuration parameters of \sext\ as follows: DETECT\_MINAREA 0.0081\,arcsec$^2$, DETECT\_THRESH 1.0, DEBLEND\_NTHRESH 64, DEBLEND\_MINCONT 0.0001, BACK\_SIZE 128, and BACK\_FILTSIZE 5. The aperture-based fluxes measured for each source are then scaled by a single factor, defined as ${f_{\rm auto, F444W}/f_{\rm aper, F444W}}$, to the total flux, where ${f_{\rm auto}}$ is the flux measured within elliptical Kron apertures with the 2.5 scaling factor. With this approach, colors remain as those measured in the aperture, whereas the fluxes used in the following analyses (for the derivation of stellar mass, $M_*$, and star formation rate, SFR) are scaled up to the total quantities. 

Once the fluxes are measured and scaled, Galactic dust reddening is corrected using the attenuation value retrieved for the coordinates of each field from NED \citep{schlegel98,schlafly11}. We adopt the canonical Milky Way dust law for the reddening curve \citep{cardelli89}.

%%%%%%%%%%%%%%%%%%%%%%%%%%%%%%%%%%
\subsection{NIRSpec MSA}\label{subsec:nirspec}
%%%%%%%%%%%%%%%%%%%%%%%%%%%%%%%%%%
\begin{figure*}[ht!]
\begin{center}
\includegraphics[scale=0.44]{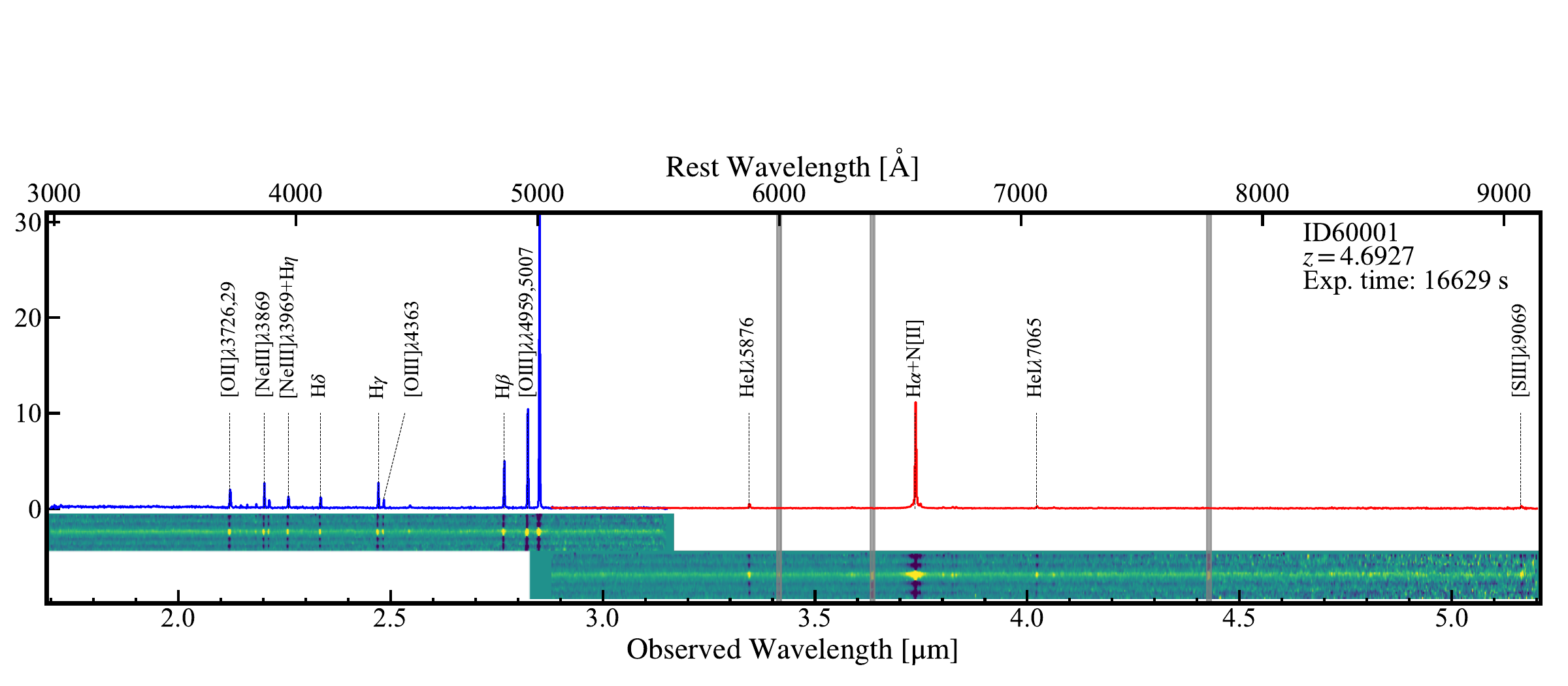}
\end{center}
\caption{Top: Reduced 1D spectra of G235M/F170LP (blue) and G395/F290LP (red) filter/grating extracted from the 2D spectra. Bottom: Reduced 2D spectra of G235M/F170LP and G395/F290LP filter/grating. Regions containing bad pixels are masked out and indicated with gray shaded areas.
}\label{fig:spec_all}
\end{figure*}
%%%%%%%%%%%%%%%%%%%%%%%%%%%%%%%%%%
We utilize the spectroscopic data of the NIRSpec Multi-Object Spectroscopy (MSA) taken as part of the program GTO~2756 \citep{stiavelli24}. The observations were conducted in the 1199par field, primarily targeting $z>2$ galaxies that were photometrically identified. Our target galaxy in this paper, ID60001, was observed for 16629\,s (excluding overhead), with each of the G235M and G395M gratings. We reduce the MSA data using {\tt msaexp}\footnote{\url{https://github.com/gbrammer/msaexp}} ({ver.~0.8.5}), following the same scheme as \citet{morishita24a}. The one-dimensional spectrum was extracted via optimal extraction. 

The reduced one-dimensional (1D) and two-dimensional (2D) spectra are displayed in Figure \ref{fig:spec_all}.

\section{ISM Analysis}\label{sec:ism}
\subsection{Emission Line Measurements} \label{sec:emission}
With the reduced NIRSpec 1D spectra, we conduct emission line fitting. For each unblended emission line, we initially fit a model with a flat continuum and one Gaussian profile. We then convolve our model with the line spread function (LSF) of \citet{isobe23b} and fit the convolved model to the observed spectrum. We obtain the best-fit model with Monte Carlo simulations, making 500 mock spectra by adding random noise to the observed flux density of each spectral pixel. The random noise is generated following a Gaussian distribution whose standard deviation is defined by the $1\sigma$ uncertainty in the observed spectrum. 
For [O{\sc iii}]$\lambda\lambda4959,5007$ and [N{\sc ii}]$\lambda\lambda6548,6584$ doublets, we fix the flux ratio to $1:2.98$ and $1:2.94$, respectively. For (partially) blended emission line complexes of [O{\sc iii}]$\lambda\lambda3727,3729$ and [N{\sc ii}]$\lambda\lambda6548,6584 + $H$\alpha$, we tie the rest-frame central wavelengths as well as line widths (FWHM) in the unit of km~s$^{-1}$ of each individual emission line. 

\begin{figure*}[ht!]
\begin{center}
\includegraphics[scale=0.48]{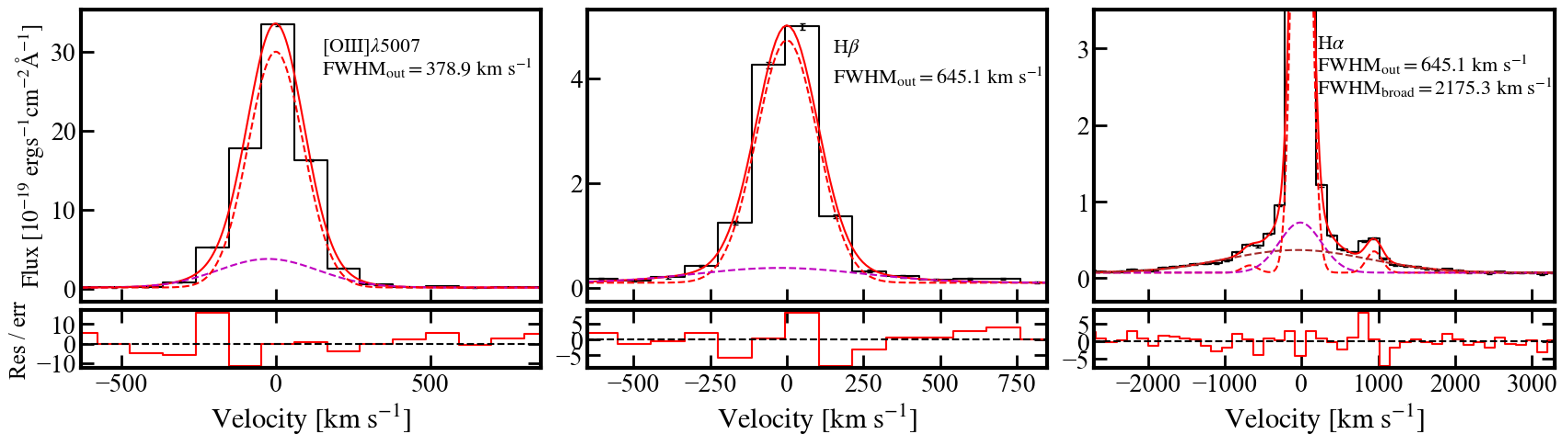}
\end{center}
\caption{Emission line fitting results for [OIII]$\lambda$5007 (left), H$\beta$ (middle), and H$\alpha$(right). The black histograms and error bars denote the NIRSpec MSA observational data. The red solid curves shows the best-fit profiles. The narrow, outflow, and broad component of each best-fit profile are displayed with red, magenta, and brown dashed curves, respectively.
}\label{fig:emlinefit}
\end{figure*}

Upon examining the residual of our initial fitting results, we find that most of the emission lines can be well-fitted with a single narrow (FWHM$_{\rm narrow}\leq300$~km~s$^{-1}$) Gaussian model with a median line width of FWHM=250.4$_{-31.7}^{+52.0}$~km~s$^{-1}$. However, for H$\beta$, [O{\sc iii}]$\lambda\lambda$4959,5007, and [N{\sc ii}]$\lambda\lambda6548,6584 + $H$\alpha$ complexes, there are significant positive, extended residuals around the wing of the emission lines, indicating potential existence of extra broad components due to outflows. We fit these emission lines again, adding a broad (FWHM$_{\rm outflow}>$FWHM$_{\rm narrow}$) Gaussian component (hereafter ``outflow component'') to each of the H$\beta$, [O{\sc iii}]$\lambda\lambda$4959, 5007, and H$\alpha$ emission lines. We tie the velocity shifts ($\Delta v$, in km~s$^{-1}$), line widths, and the flux ratio of the outflow to narrow components ($f_{\rm o/n}$) between the [O{\sc iii}]$\lambda\lambda$4959, 5007 doublet as well as between H$\beta$ and H$\alpha$, respectively. We then repeat the fitting process, obtaining the best-fit model and parameters. For H$\beta$ and [O{\sc iii}]$\lambda\lambda$4959, 5007 doublet, the double Gaussian model is preferred over the single Gaussian one in terms of improved fitting residual (Figure \ref{fig:emlinefit}) and decreased Bayesian information criterion (BIC).
%, and c) the reasonable best-fit parameters in \textcolor{red}{FWHM$_{\rm narrow} = xxx (yyy)$, FWHM$_{\rm outflow} = xxx (yyy)$, $f_{\rm o/n} = xxx (yyy)$} for H$\beta$ ([O{\sc iii}]$\lambda$5007).
%The goodness of fit of the single and double Gaussian model is compared using the Bayesian information criterion \citep[$\Delta$BIC;][]{schwarz78}, which is given by $\Delta\mathrm{BIC} = \chi^2_2 - \chi^2_1 + (k_2-k_1)\log N$. Here $\chi^2_1$($\chi^2_2$) and $k_1$($k_2$) refers to the $\chi^2$ and number of free parameters for single (double) Gaussian model, respectively. We obtain $\Delta$BIC $=XX.X$, suggesting that the outflow model better describes the observed data.
For H$\alpha$, we observe that the double Gaussian model still results in significant positive residuals around the wing of the emission lines. We therefore add a third broad Gaussian component with FWHM$_{\rm broad}>$FWHM$_{\rm out}$ to H$\alpha$ and repeat the fitting procedure mentioned above. Our fitting results imply that the H$\alpha$ emission line is best described by the triple Gaussian model with FWHM$_{\rm broad}= 2175_{-81}^{+133}$~km~s$^{-1}$, FWHM$_{\rm out}= 645_{-49}^{+72}$~km~s$^{-1}$, FWHM$_{\rm narrow}= 230_{-2}^{+2}$~km~s$^{-1}$. The best-fit spectra of [N{\sc ii}]$\lambda\lambda6548,6584 + $H$\alpha$ is shown in Figure \ref{fig:emlinefit}.
%The best-fit FWHM and outflow velocity are similar to those from typical stellar outflow (\textcolor{red}{References?}). 

We summarize the FWHM, flux, and rest-frame equivalent width (EW$_0$) obtained from the best-fit model of each identified emission line in Table \ref{tab:lines}. For the following analyses, we focus on the ISM without outflow gas, adopting the fluxes of the narrow components of H$\beta$, [O{\sc iii}]$\lambda\lambda$4959, 5007, and H$\alpha$. For the other emission lines where potential outflow components are undetected due to limited S/N, we adopt the fluxes estimated from the single Gaussian fit results.

%%%%%%%%%%%%%%%%%%%%%%%%%%%%%%%%%%
%%%%%%%%%%%%%%%%%%%%%%%%%%%%%%%%%%

\begin{deluxetable}{lcc}[h]
\tablecaption{Emission line measurements \label{tab:lines}}
\tablewidth{0pt}
\tablehead{
\colhead{Line} &
\colhead{Flux} &
\colhead{EW$_0$} \\
 & ($10^{-19}$~erg~s$^{-1}$~cm$^{-3}$) & (\AA)
}
\startdata
[O{\sc ii}]$\lambda\lambda$3726,3729\tablenotemark{a} & $46.5^{+0.8}_{-0.8}$ & $73.0^{+2.4}_{-2.2}$ \\ 
%H12 & $8.55^{+0.18}_{-1.03}$ & $0.65^{+0.01}_{-0.01}$ \\ 
%H11 & $4.8^{+0.5}_{-0.5}$ & $0.65^{+0.01}_{-0.01}$ \\
H$\theta$ & $3.4^{+0.4}_{-0.4}$ & $5.8^{+0.7}_{-0.7}$ \\
H$\eta$ & $4.7^{+0.4}_{-0.4}$ & $8.2_{-0.8}^{+0.8}$ \\
{[}Ne{\sc iii}]$\lambda$3869 & $39.1^{+0.6}_{-0.6}$ & $77.9^{+3.5}_{-3.2}$ \\ 
He{\sc i}$\lambda$3889+H$\zeta$\tablenotemark{a} & $13.8^{+0.5}_{-0.5}$ & $27.7_{-1.5}^{+1.6}$ \\
{[}Ne{\sc iii}]$\lambda$3969+H$\epsilon$\tablenotemark{a} & $26.4_{-0.5}^{+0.6}$ & $48.0^{+2.3}_{-2.1}$ \\ 
He{\sc i}$\lambda$4026 & $1.8^{+0.3}_{-0.3}$ & $4.0^{+0.7}_{-0.6}$ \\ 
H$\delta$ & $22.8^{+0.5}_{-0.5}$ & $51.3^{+2.4}_{-2.2}$ \\ 
H$\gamma$ & $44.7^{+0.5}_{-0.6}$ & $111.5^{+4.8}_{-4.4}$ \\ 
{[}O{\sc iii}]$\lambda$4363 & $13.5^{+0.4}_{-0.4}$ & $33.5^{+1.9}_{-1.7}$ \\ 
He{\sc i}$\lambda$4388 & $1.2^{+0.3}_{-0.3}$ & $3.5_{-1.0}^{+0.9}$ \\
He{\sc i}$\lambda$4471 & $5.8^{+0.4}_{-0.4}$ & $12.4_{-1.0}^{+1.1}$ \\
He{\sc ii}$\lambda$4686 & $2.5^{+0.4}_{-0.4}$ & $6.7_{-1.0}^{+1.1}$ \\
H$\beta$ (narrow) & $82.3^{-1.2}_{+1.1}$ & $191.7^{+8.3}_{-6.8}$ \\ 
H$\beta$ (outflow) & $13.7^{+0.8}_{-0.7}$ & $31.9^{+2.4}_{-2.1}$ \\ 
{[}O{\sc iii}]$\lambda$4959 (narrow) & $178.6^{+1.3}_{-1.3}$ & $303.8^{+13.4}_{-12.8}$ \\ 
{[}O{\sc iii}]$\lambda$4959 (outflow) & $21.5^{+1.3}_{-1.3}$ & $36.6^{+3.4}_{-3.2}$ \\ 
{[}O{\sc iii}]$\lambda$5007 (narrow) & $532.5^{+3.9}_{-3.9}$ & $905.4^{+39.9}_{-38.2}$ \\ 
{[}O{\sc iii}]$\lambda$5007 (outflow) & $64.2^{+3.8}_{-3.8}$ & $109.2^{+10.2}_{-9.6}$ \\ 
{[}Fe{\sc ii}]$\lambda$5158 & $1.3^{+0.3}_{-0.3}$ & $4.2^{+1.0}_{-0.9}$ \\ 
{[}Fe{\sc iii}]$\lambda$5270+{[}Fe{\sc ii}]$\lambda$5273\tablenotemark{a} & $4.2^{+0.6}_{-0.6}$ & $13.7^{+2.3}_{-2.1}$ \\ 
%{[}O{\sc i}]$\lambda$5576 & $0.7^{+0.3}_{-0.3}$ & $0.65^{+0.01}_{-0.01}$ \\ 
He{\sc i}$\lambda$5876 & $13.4^{+0.4}_{-0.3}$ & $42.1_{-1.8}^{+1.9}$ \\
{[}O{\sc i}]$\lambda$6046 & $0.8^{+0.2}_{-0.2}$ & $3.0^{+0.8}_{-0.8}$ \\ 
{[}O{\sc i}]$\lambda$6300 & $2.8^{+0.2}_{-0.2}$ & $10.7^{+1.1}_{-1.0}$ \\ 
{[}S{\sc iii}]$\lambda$6312 & $1.6^{+0.2}_{-0.2}$ & $6.2^{+0.9}_{-0.9}$ \\ 
{[}N{\sc ii}]$\lambda$6548 & $2.3^{+0.2}_{-0.2}$ & $7.2^{+0.6}_{-0.5}$ \\ 
H$\alpha$ (narrow) & $257.5^{+4.4}_{-3.7}$ & $800.0^{+51.5}_{-31.7}$ \\ 
H$\alpha$ (outflow) & $42.9^{+2.8}_{-2.2}$ & $133.7^{+11.5}_{-8.8}$ \\ 
H$\alpha$ (broad) & $65.2^{+2.7}_{-3.5}$ & $203.9^{+11.3}_{-10.6}$ \\ 
{[}N{\sc ii}]$\lambda$6584 & $6.8^{+0.4}_{-0.4}$ & $21.1^{+1.7}_{-1.4}$ \\ 
He{\sc i}$\lambda$6678 & $2.7^{+0.3}_{-0.2}$ & $9.0_{-0.9}^{+1.1}$ \\
{[}S{\sc ii}]$\lambda$6716 & $3.6^{+0.2}_{-0.2}$ & $13.0^{+1.2}_{-1.1}$ \\ 
{[}S{\sc ii}]$\lambda$6731 & $3.0^{+0.2}_{-0.2}$ & $10.9^{+1.0}_{-1.0}$ \\ 
He{\sc i}$\lambda$7065 & $7.0^{+0.3}_{-0.3}$ & $32.0^{+1.8}_{-1.7}$ \\
{[}Ar{\sc iii}]$\lambda$7136 & $3.6^{+0.3}_{-0.3}$ & $16.5^{+1.7}_{-1.5}$ \\ 
{[}O{\sc ii}]$\lambda$7319 & $1.4^{+0.2}_{-0.2}$ & $5.3^{+0.7}_{-0.8}$ \\ 
{[}O{\sc i}]$\lambda$8446 & $3.1^{+0.5}_{-0.5}$ & $12.8^{+2.5}_{-2.2}$ \\ 
{[}S{\sc iii}]$\lambda$9069 & $9.3^{+0.8}_{-0.7}$ & $57.2^{+7.2}_{-6.3}$ \\ 
\hline
\enddata
\tablenotetext{a}{Blended emission lines.}
\end{deluxetable}
%%%%%%%%%%%%%%%%%%%%%%%%%%%%%%%%%

\subsection{On the Existence of AGN\label{subsec:agn}}
We first examine the possible existence of AGN activity in ID60001. As discussed in Section \ref{sec:emission}, we find a relatively broad component (FWHM=$2175_{-81}^{+133}$~km~s$^{-1}$) in the H$\alpha$ emission line in addition to the outflow component. Such a broadening in Balmer lines is sometimes considered to originate from the broad-line region (BLR) and hence used as evidence of AGN activities especially at high-$z$ \citep[e.g.][]{harikane23,larson23,maiolino24}. However, other nontypical stellar explosions/eruptions such as Wolf-Rayet stars, tidal disruption events (TDEs), and type IIn supernovae (SNe) can also result in long-lasting, broad components in the Balmer emission lines \citep[e.g.][]{kokubo24}. The broadening of H$\alpha$ alone hence cannot serve as decisive evidence of AGN activity in ID60001. 
%\textcolor{red}{Source of broad components: shocks, etc.}

\begin{figure*}[ht!]
\begin{center}
\includegraphics[scale=0.6]{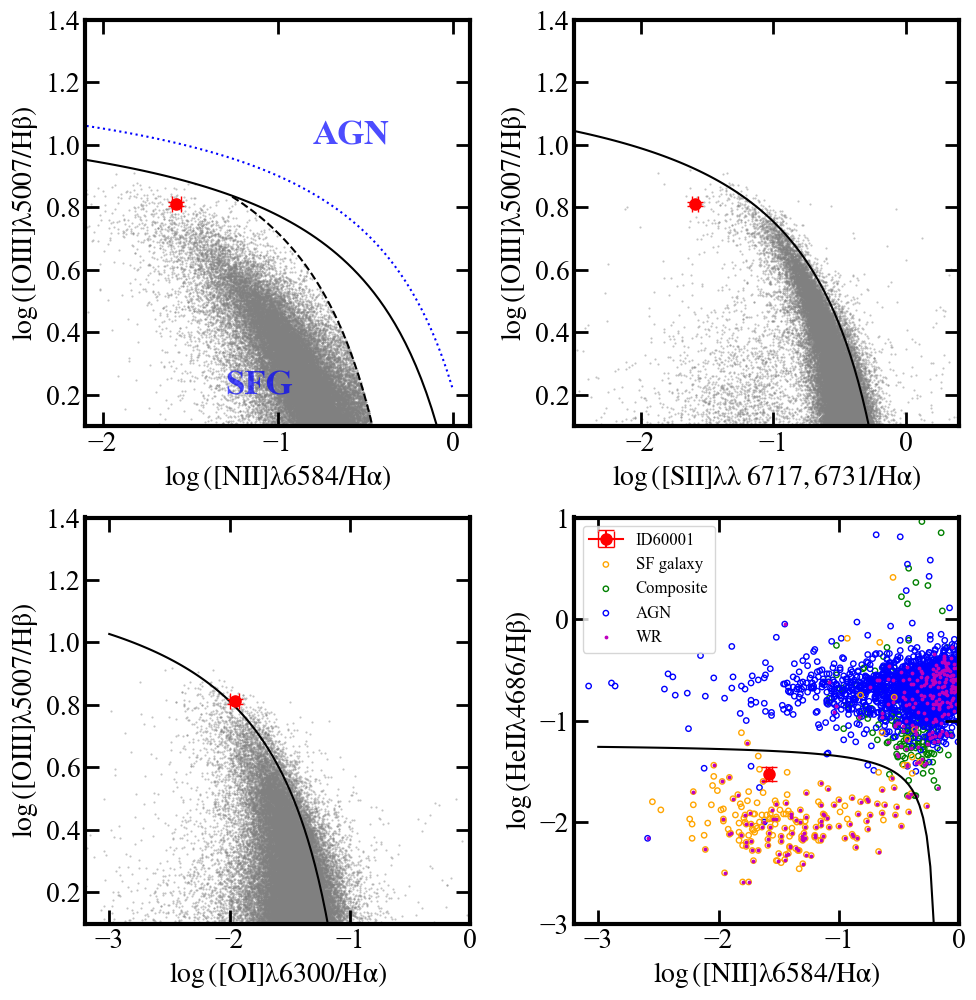}
\end{center}
\caption{Optical emission line diagnostics for AGN/SFG separation. Top-left: BPT diagram where the red data point indicate ID60001. The grey circles represent the SF galaxies taken from SDSS-DR8 \citep{kauffmann03,brinchmann04,tremonti04}. The black solid and dashed curve indicate the AGN-SF galaxy separation derived by \citet{kewley01} and \citet{kauffmann03}, respectively. The blue dotted curve is the AGN-SF separation at $z\sim3$ from \citet{kewley13}. Top-right and Bottom-left: Same as top-left, but for VO87-S{\sc ii} and VO87-O{\sc i} diagram, respectively. Bottom-right: He2-N2 diagram where the black solid curve indicate the AGN-SF galaxy separation obtained by \citet{shirazi12}. The small orange, green, and blue open circles denote the SF galaxy, composite galaxy, and AGN classified by \citet{shirazi12}. Objects with WR features are shown with small magenta circles. In all four diagnostic plots, ID60001 locates in the SF galaxy regime and overlaps with other SF galaxies in literature.}\label{fig:bpt}
\end{figure*}

To further investigate whether or not AGN contributes to the ISM ionization of ID60001, we compare the emission line ratios with different optical emission line diagnostics such as BPT ([O{\sc iii}]/H$\beta$ vs. [N{\sc ii}]/H$\alpha$; \citealp{kewley01}), VO87-S{\sc ii} (O{\sc iii}/H$\beta$ vs. [S{\sc ii}]/H$\alpha$; \citealp{vo87}), VO87-{\sc OI} (O{\sc iii}/H$\beta$ vs. [O{\sc i}]/H$\alpha$), He2-N2 (He{\sc ii}/H$\beta$ vs. [N{\sc ii}]/H$\alpha$) diagrams. As shown in Figure \ref{fig:bpt}, our object is located in the SF region of all four diagnostics except for the VO87-{\sc OI} diagram, where our object sits on the edge of SF-AGN boundary. Although He{\sc ii}$\lambda 4686$ with a high ionization potential of 54.4~eV is detected, its relative intensity to H$\beta$ is consistent with the SF scenario in the He2-N2 diagram. The moderate EW$_0$ of He{\sc ii}$\lambda 4686$ EW$_0$(He{\sc ii}$\lambda 4686$)=6.7~\AA\ is also comparable to the typical cases observed in prominent SF galaxies\citep[e.g.,][]{guseva00,martin23}, indicating that the hard ionizing photons producing the He{\sc ii}$\lambda 4686$ emission in ID60001 is likely attributed to extreme SF activities that sometimes feature WR stars \citep[i.e.][]{shirazi12,morishita24a}. Although further multi-wavelength observational data (i.e. X-ray, rest-frame UV and mid-IR) is needed to conclude whether or not the ionizing source of ID60001 is an AGN, our available evidence based on rest-frame optical emission line diagnostics strongly
supports the scenario that ID60001 is a SF galaxy without AGN activity.
%%%%%%%%%%%%%%%%%%%%%%%%%%%%%%%%%%
%%%%%%%%%%%%%%%%%%%%%%%%%%%%%%%%%%

\subsection{Dust Extinction}\label{subsec:extinct}
We correct for the dust extinction in ISM with isolated, clearly detected Hydrogen recombination lines in the NIRSpec spectra, i.e. H$\alpha$, H$\beta$, H$\gamma$, and H$\delta$. The intrinsic emissivity of each line is calculated based on case B recombination with electron temperature $T_{\rm e} = 15000$~K. We find this assumption to be resonable based on the analyses in the following section. We derive attenuation, $A(V)$, by fitting the observed line ratios with respect to H$\beta$, $R_\lambda = I_\lambda / I_{\mathrm{H}\beta}$, to those predicted by different extinction laws
in SMC \citep{gordon03} and \citet{calzetti00}. For both models, we find $A(V)=0.0\ $yields the best-fit results. We notice that our best-fit dust extinction correction for ID60001 is different from \citet{stiavelli24}, who obtained $A(V)=1.09$ using the total fluxes (narrow+broad+outflow) of H$\alpha$. For other Balmer lines, the broad and/or outflow components are unresolved due to lower S/N while the FWHMs of the single Gaussian profile are consistent with the narrlow H$\alpha$ component. To ensure the consistency of the analyses, we hence adopt $A(V)=0.0$ (i.e. no extinction) for ID60001. 

\subsection{Electron Temperature and Density}\label{subsec:te_ne}
We derive electron density ($n_e$) and electron temperature ($T_e$) based on emission line ratios. Since $n_e$ has a more significant dependency on $T_e$ than $T_e$ on $n_e$, we first evaluate $T_e$ of the O$^{2+}$ zone ($T_e$([O{\sc iii}])) using the [O{\sc iii}]$\lambda\lambda$4363/5007 flux ratio. With the \texttt{getTemDen} task from PyNeb \citep{pyneb15} we obtain $T_e$([O{\sc iii}])$=(1.69\pm0.03)\times 10^4$~K under the assumption of $n_e=100$~cm$^{-3}$. Note that adopting $n_e=1000$~cm$^{-3}$ will results in a $<1\%$ change in $T_e$([O{\sc iii}]) that is well within the errors. From $T_e$([O{\sc iii}]) we extrapolate $T_e$ of the O$^{+}$ zone ($T_e$([O{\sc ii}])) using the prescription of \citet{izotov06}. We then estimate $n_e$ with the [S{\sc ii}]$\lambda\lambda$6716/6731 doublet line ratio and $T_e$([O{\sc iii}]), obtaining the result of $n_e=360_{-173}^{+177}$~cm$^{-3}$. We do not use the [O{\sc ii}]$\lambda\lambda$3726,3729 doublet for $n_e$ estimation because their peaks are fully blended in the G235M spectrum and may cause large uncertainties in line ratio measurements. Our measured $T_e$([O{\sc iii}]) and $n_e$ values are generally consistent with the mean values at the similar redshift \citep[e.g.][]{isobe23b}.

%%%%%%%%%%%%%%%%%%%%%%%%%%%%%%%%%%%%
\begin{table}[ht!]
\centering
\caption{ISM and host Properties} \label{tab:prop}
\begin{tabular}{lc}
%\tablewidth{0pt}
\hline
\hline
\multicolumn{2}{c}{ISM properties} \\ 
%Parameters & Extinction corrected & No extinction \\
\hline
%\decimals
$A_V$(ISM) & 0.0 \\
$A_V$(SED) & $0.13\pm0.03$ \\
$\log U$(SED) & $-1.69_{-0.23}^{+0.14}$ \\$T_e$([O{\sc iii}]) & $(1.61\pm0.04)\times 10^4$~K \\
%$T_e$([O{\sc iii}]) w/ outflow & $(1.26\pm0.02)\times 10^4$~K \\
$n_e$([S{\sc ii}]) & $345_{-172}^{+184}$~cm$^{-3}$ \\
%$n_e$([S{\sc ii}]) w/ outflow & $359_{-211}^{+206}$~cm$^{-3}$ \\
$12+\log({\rm O/H})$ & $7.75\pm 0.03$ \\
%$12+\log({\rm N/H})$ & $6.84\pm 0.03$ \\
$\log({\rm N^+/O^+})$ & $-0.85\pm 0.03$ \\
$\log({\rm ICF(N^+/O^+)})$ & 0.09 \\
$\log({\rm N/O})$ & $-0.76\pm 0.03$ \\
%$12+\log({\rm Ne/H})$ & $6.99\pm 0.03$ \\
%$12+\log({\rm S/H})$ & $6.07_{-0.08}^{+0.06}$ \\
%$12+\log({\rm Ar/H})$ & $5.38\pm 0.04$ \\
\hline
\multicolumn{2}{c}{Host properties} \\
\hline
$\log(M_*/M_\odot)$ & $9.34_{-0.10}^{+0.08}$ \\
SFR$_{\rm 10Myr}$ (SED) & $13.49_{-2.33}^{+1.94}~M_\odot~{\rm yr}^{-1}$ \\
SFR (H$\alpha$) & $21.72_{-0.31}^{+0.37}~M_\odot~{\rm yr}^{-1}$ \\
%$\log ({\rm Age/[yr]})$ & $6.75_{-0.04}^{+0.29}$ \\
%$Av$(SED) & $0.7_{-0.01}^{+0.01}$ \\
\hline
\end{tabular}
\end{table}
%%%%%%%%%%%%%%%%%%%%%%%%%%%%%%%%%%%%%%

\subsection{Chemical Abundances \label{subsec:abundance}}
We derive the chemical abundances O/H and N/O of ID60001 based on $T_e$ and $n_e$ derived in the previous subsection and the extinction-corrected emission line fluxes. We first estimate ion abundances with emission line flux ratios relative to H$\beta$ using the \texttt{getIonAbundance} package of PyNeb, then converting ion abundances to the corresponding element abundances using ionization correction factors (ICFs).

For oxygen abundance O/H, we estimate O$^+$/H$^+$ and O$^{2+}$/H$^+$ with [O{\sc ii}]$\lambda\lambda3726+3729$ and [O{\sc iii}]$\lambda5007$, respectively. We then derive O/H with the ICF of Equation 17 in \citet{izotov06}, which takes into account the contribution of O$^{3+}$ ion from high excitation H{\sc ii} region traced by He{\sc ii}$\lambda4686$. The He$^+$/H and He$^{2+}$/H ion abundances are estimated with He{\sc i}$\lambda5876$ and He{\sc ii}$\lambda4686$, respectively.

For N/O, we estimate N$^+$/H$^+$ with the rest-frame optical emission line [N{\sc ii}]$\lambda6548$, and directly calculate N/O with the ICF in \citet{amayo21}:
\begin{equation}
    \rm{\frac{N}{O} = ICF(\frac{N^+}{O^+})\frac{N^+/H^+}{O^+/H^+}}
\end{equation}

\begin{figure*}[ht!]
\begin{center}
\includegraphics[scale=0.47]{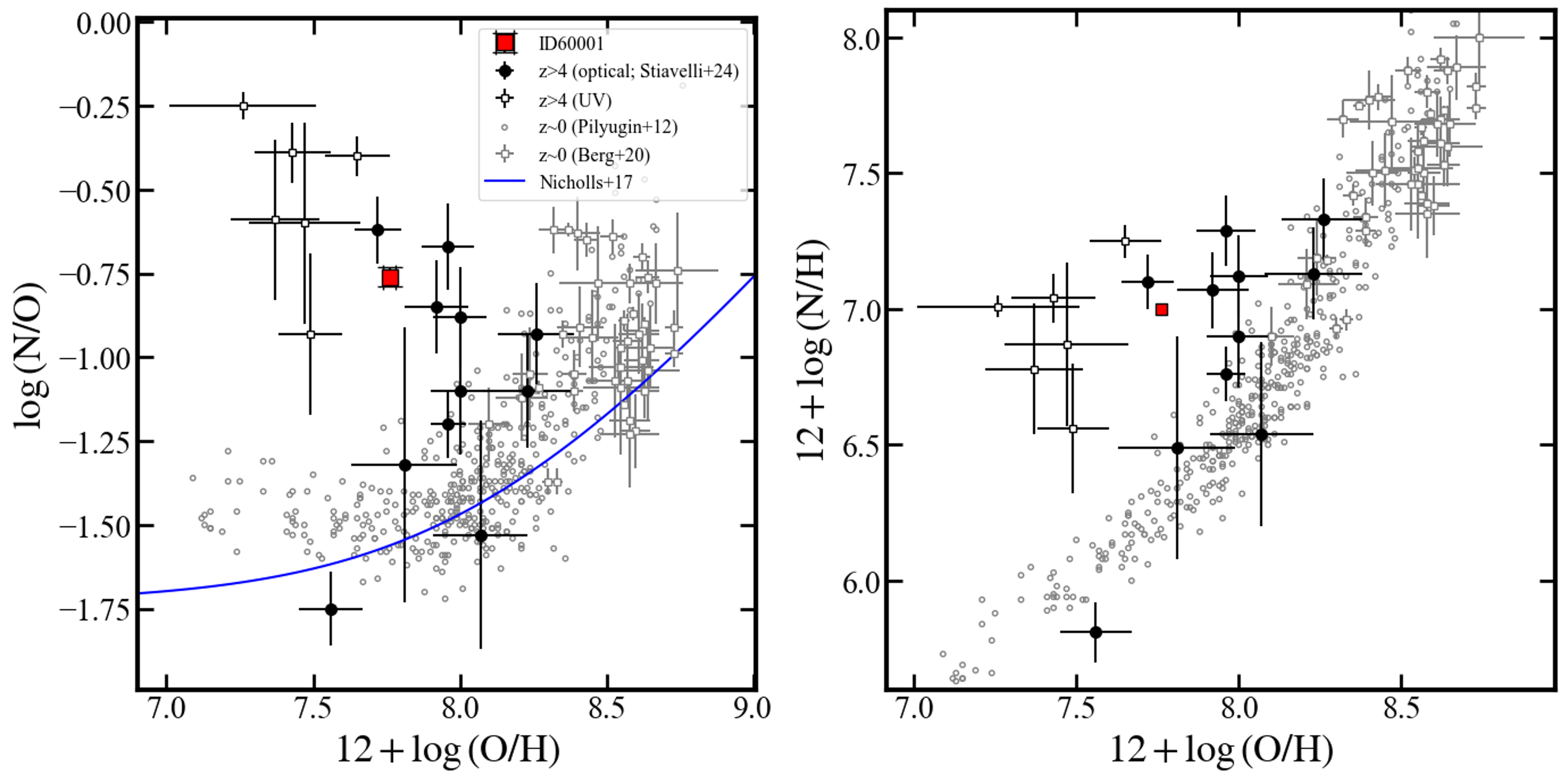}
\end{center}
\caption{Left: N/O abundance ratio versus gas phase metallicity of ID60001 (red square) and other high-$z$ samples based on optical (solid black circles) and UV (open black circles) emission lines. For comparison, local samples \citep{pilyugin12,berg20} and relation \citep{nicholls} are also displayed. Right: N/H abundance versus gas phase metallicity. Symbols are identical to the left panel.}\label{fig:n_o}
\end{figure*}

We obtain 12+log(O/H) = 7.75$\pm$0.01 and log(N/O) = -0.76$\pm$0.03. We display our results in Figure \ref{fig:n_o} along with other $z>4$ \citep{stiavelli24} and local \citep{pilyugin12} results based on the optical [N{\sc ii}]$\lambda6548$ emission line. Compared with the local sample, ID60001 has significant nitrogen overabundance with respect to its metallicity. Such an overabundance is comparable with other ``N-rich" galaxies at $z>4$ based on the high-ionization UV emission lines from previous JWST studies \citep[e.g.][]{isobe23b,schaerer24,ji24,topping24}. 

Apart from N/O abundance, we also derive N/H abundance in Figure \ref{fig:n_o} assuming N/H = N/O $\times$ O/H. Similar to N/O, ID60001 has an elevated N/H value from local objects at the corresponding metallicity, confirming that nitrogen is enriched compared to both oxygen and hydrogen. In Section \ref{subsec:discuss_no}, we will discuss the origins of nitrogen overabundance for ID60001 in detail.

%%%%%%%%%%%%%%%%%%%%%%%%%%%%%%%%%%
%%%%%%%%%%%%%%%%%%%%%%%%%%%%%%%%%%

\section{Host Galaxy Analysis}\label{sec:host}
\subsection{SED fitting and galaxy properties}\label{subsec:sed}
%We conduct SED fitting for ID60001 with Bagpipe, simultaneously fitting the NIRSpec MSA spectra and NIRCam photometry with \texttt{Bagpipe} \citep{carnall18,carnall19}. The model galaxy spectra are generated with the 2016 version of \citet{bc03} model and the default \citet{kroupa02} initial mass function (IMF). We adopt the a non-parametric continiuity SF history \citep{leja19} with five bins between 0, 10, 100, 250, 500, and 1000~Myr of lookback time, where the SFR in each bin is constant. Throughout the fitting, we assume the \citet{calzetti00} dust extinction law and vary the redshift witin a small range ($\pm0.001$) around the $z_{\rm spec}$. Our best-fit SED and SF history is shown in Figure \ref{fig:sed}. With our SED analyses, we obtain the best-fit result of $\log (M_*/M_\odot)=8.68_{-0.01}^{+0.01}$, SFR $=5.05_{-0.08}^{+0.08}~M_\odot~{\rm yr}^{-1}$ featuring a bursty SF history within the lookback time of 10~Myr.
We conduct SED fitting for ID60001 with \texttt{PROSPECTOR} \citep{prospect}, utilizing NIRCam multiband photometry (Section \ref{subsec:photom}). The model spectra are generated from the Flexible Stellar Population Synthesis \citep[FSPS;][]{conroy09,cg10} package with the Modules for Experiments in Stellar Astrophysics Isochrones and Stellar Tracks \citep[MIST;][]{choi16}, where the boost of the ionizing flux production of massive stars is included \citep{choi17}. We assume the \citet{chab03} initial mass function (IMF), the SMC dust extinction law \citep{gordon03}, and the \citet{madau95} intergalactic medium (IGM) attenuation model. We adopt the a non-parametric continuity SF history \citep{leja19} with five bins between 0, 10, 100, 250, 500, and 1000~Myr of lookback time, where the SFR in each bin is constant. Throughout the SED analysis, we exploit the ISM properties derived from spectroscopic data in Section \ref{sec:ism}, fixing the redshift to the spectroscopically determined one and setting the priors of $A_V$, ionization parameter ($\log U$), and metallicity $\log(Z/Z_\odot)$ to be centering at the values inferred by the spectroscopic analysis; i.e. a clipped Gaussian distribution $A_V=0.0$ with a dispersion of 0.2 and the range of $0.0\leq A_V\leq0.5$, a clipped Gaussian distribution centering at $\log(U)=-2.3$ with a dispersion of 0.5 and the range of $-3.0\leq\log(U)\leq-1.5$, and a Gaussian distribution centering at $\log(Z/Z_\odot)=-0.95$ with a dispersion of 0.5, respectively. The prior of $M_*$ is set to be a uniform distribution at $6.0\leq\log(M_*/M_\odot)\leq12.0$. 

We search for the best-fit SED model and the corresponding parameters using the Markov Chain Monte Carlo method using \texttt{EMCEE}, obtaining the best-fit result of $\log (M_*/M_\odot)=9.34_{-0.10}^{+0.08}$ and SFR$=13.49_{-2.23}^{+1.94}~M_\odot~{\rm yr}^{-1}$. The SFR from SED fitting is represented by the values in the most recent SFH bin of a $0-10$~Myr lookback time, which is slightly smaller than the SFR derived from H$\alpha$ emission line luminosity, SFR(H$\alpha$)$=21.72_{-0.31}^{+0.37}~M_\odot~{\rm yr}^{-1}$. The measured properties from SED fitting are summarized in Table \ref{tab:prop}.

%%%%%%%%%%%%%%%%%%%%%%%%%%%%%%%%%%
%\begin{figure*}[ht!]
%\begin{center}
%\includegraphics[scale=0.8]{sed_results.png}
%\end{center}
%\caption{Best-fit SED of ID60001 (orange curve) with Bagpipe. The observed photometry is indicated with blue circle. The blue and red curves show the observed NIRSpec MSA data. In the small panel we display the best-fit SF history with errors indicated by the grey shaded regions.}\label{fig:sed}
%\end{figure*}
%%%%%%%%%%%%%%%%%%%%%%%%%%%%%%%%%%

\subsection{Galaxy Morphology}\label{subsec:morph}
%%%%%%%%%%%%%%%%%%%%%%%%%%%%%%%%%%
\begin{figure*}[htb!]
\begin{center}
\includegraphics[scale=0.55]{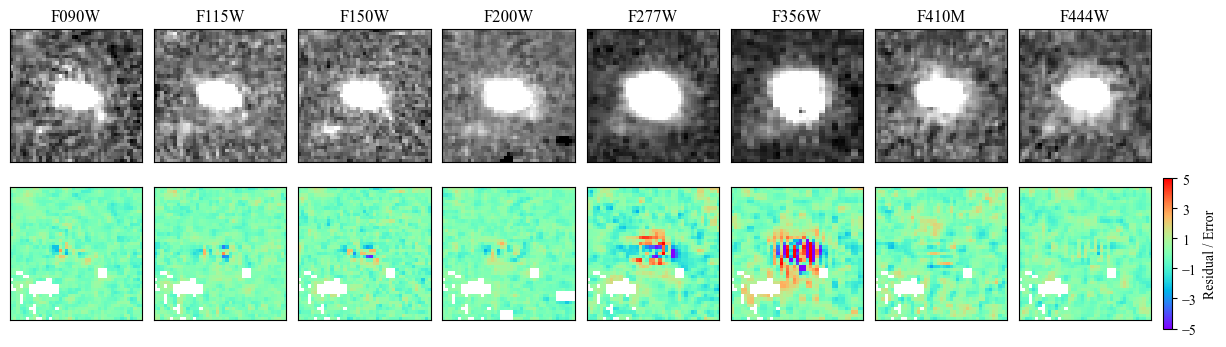}
\end{center}
\caption{Galfit fitting results for ID60001. Top and bottom row shows the observed data and residual from best-fit models, respectively. Our results show clumpy residuals only in F277W and F356W filters where H$\beta$+[O{\sc iii}] and H$\alpha$ are included, respectively, indicating turbulent features from ionized gas.
}\label{fig:galfit}
\end{figure*}
%%%%%%%%%%%%%%%%%%%%%%%%%%%%%%%%%%

We conduct two-dimensional profile fitting with Galfit on broadband images of ID60001. Specifically, we first fit a separate model to each of the eight filter images except for F277W and F356W where strong emission lines of H$\beta$+[O{\sc iii}]$\lambda\lambda4959,5007$ and H$\alpha$ are included. For each filter, we construct the empirical point spread function (PSF) model by selecting and stacking bright stars in the same field. We then fit the images with different models, finding that for all of the F090W, F115W, F150W, F200W, F410M, and F444W filter images, the best-fit model is described by two Sersic profiles. Assuming the spatial distributions of stellar population do not vary with wavelengths, we fix the Sersic index ($n$), position angle (P.A.), and axis ratio ($a/b$) to the median values of the best-fit results in the five filter images, and rerun the fitting to all eight filter images. With such an approach, we obtain the overall best-fit results of the two Sersic components separated by 0.85~kpc with $n_1=5.64$, effective radius $r_{e1}({\rm F090W}) = 0.60$~kpc (rest-frame UV) %, $r_e1({\rm F200W}) = 0.43$~kpc (rest-frame optical), $r_e1({\rm F444W}) = 0.15$~kpc (rest-frame IR) 
for component 1, and $n_2=0.77$, $r_{e2}({\rm F090W}) = 0.29$~kpc %, $r_e2({\rm F200W}) = 0.43$~kpc, $r_e2({\rm F444W}) = 0.15$~kpc
for component 2. The residuals of our best-fit results is shown in Figure \ref{fig:galfit}. In F277W and F356W image covering H$\beta$+[O{\sc iii}] and H$\alpha$ emission lines, respectively, we find large clumpy residuals that is absent in the other filter images, indicating disturbed features from ionized gas that are likely associated with stellar wind.
%%%%%%%%%%%%%%%%%%%%%%%%%%%%%%%%%%
\begin{figure}[htb!]
\begin{center}
\includegraphics[scale=0.41]{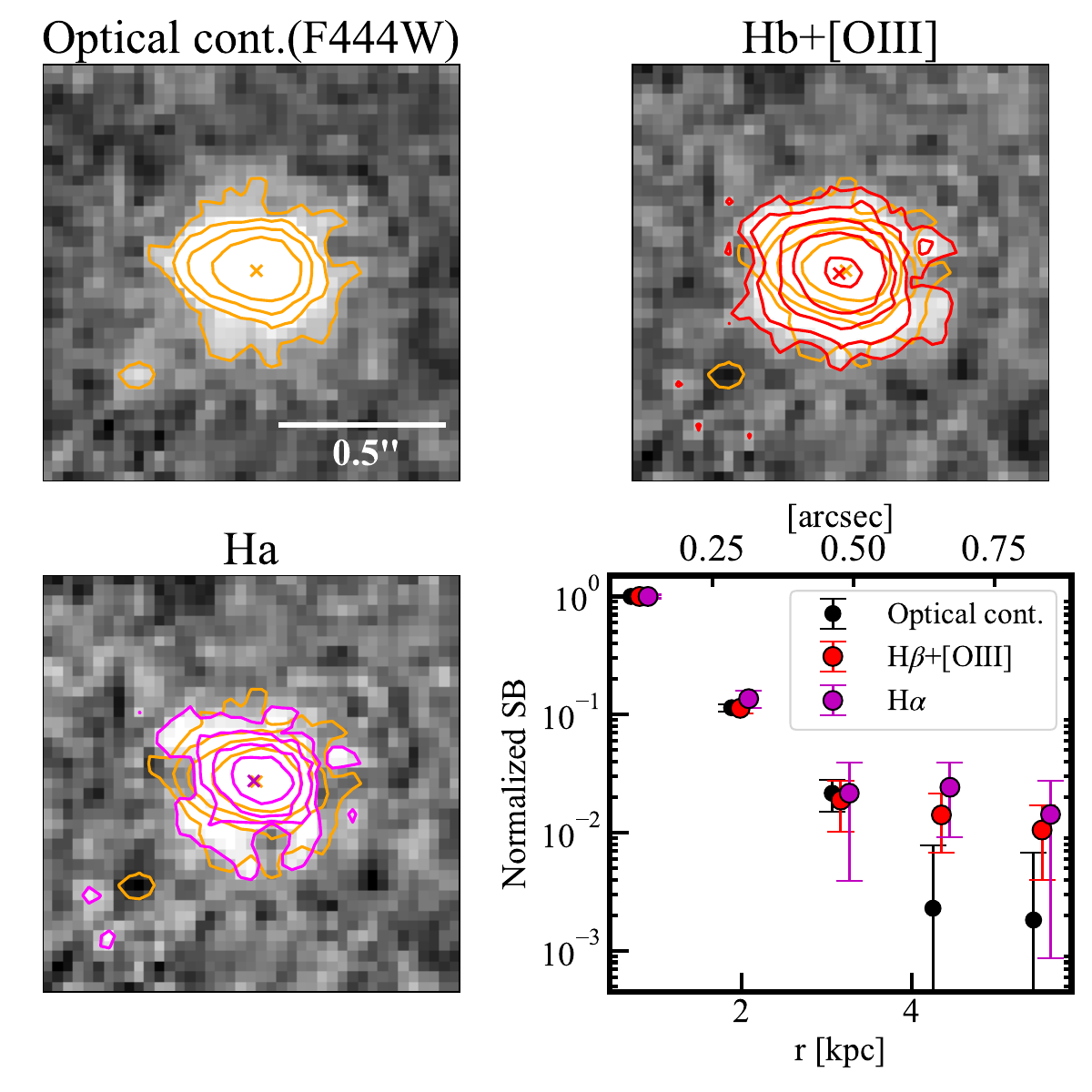}
\end{center}
\caption{The continuum image, H$\beta$+[O{\sc iii}] and H$\alpha$ emission line maps, and their radial profiles. For continuum and emission line images, the crossmarks indicate the position of galaxy centroid and the contours denote (2.5, 5, 10, 20)$\sigma$ levels in each corresponding image.
}\label{fig:emsb}
\end{figure}
%%%%%%%%%%%%%%%%%%%%%%%%%%%%%%%%%%

We further investigate the spatial distribution of H$\beta$+[O{\sc iii}]$\lambda\lambda4959,5007$ and H$\alpha$ emission by constructing and analyzing the emission line map based on broadband images in the same manner as \citet{zhang24}. Here we use F444W image as the continuum image and subtract it from F277W and F356W images to obtain the H$\beta$+[O{\sc iii}] and H$\alpha$ emission line map, respectively. Before subtraction, we match the PSF of each filter image to that of the F444W image. We do not perform continuum correction during the subtraction because our best-fit SED shows a flat continuum in the rest-frame optical-IR range. As shown in Figure \ref{fig:emsb}, both H$\beta$+[O{\sc iii}] and H$\alpha$ show spatial extension beyond the continuum with little offset in centroid. In Figure \ref{fig:emsb} we also analyze the radial profiles of H$\beta$+[O{\sc iii}], H$\alpha$ emission line maps and the continuum image. Our results show that both H$\beta$+[O{\sc iii}], H$\alpha$ emission are spatially extended beyond the continuum at $\sim0.6$~kpc from the center. Our morphological analyses, together with the broad outflow components detected in H$\beta$, [O{\sc iii}] and H$\alpha$
emission line profiles from NIRSpec MSA spectra (Section \ref{sec:emission}), indicate that ID60001 exhibits gas turbulence and outflow originated from stellar winds.
%\textcolor{blue}{We make emission line images.}
\section{Discussions}\label{sec:discuss}

\subsection{Origin of N/O enhancement}\label{subsec:discuss_no}
We examine the three possible scenarios for elevated N/O abundance discussed by \citet{stiavelli24}: i) infall of pristine gas diluting all metal element abundances but not the ratios; ii) nitrogen enrichment by WR stars; iii) Type II SN winds reducing oxygen abundance before intermediate-mass stars enrich nitrogen. 

For scenario i), the infall of pristine gas %and scenario iii), Oxygen dilution through SN winds,
would result in the dilution of the abundance ratios of heavy elements with respect to Hydrogen, resulting in an O/H abundance deficit at the given $M_*$. We compare the measured O/H abundance and $M_*$ of ID60001 with the mass-metallicity relation (MZR) of \citet{morishita24b}, finding that ID60001 is generally in line with the MZR at $z=3-10$ despite a metallicity deficit of $\Delta\log({\rm O/H})\sim-0.1$~dex, which is within the 0.2-0.3~dex scatter of MZR \citep{morishita24b}. We thus do not find evidence for significant dilution from pristine gas infall from the comparison with the MZR. The infall of pristine gas would also cause the galaxy to move horizontally towards the negative $12+\log({\rm O/H})$ direction in the left panel of Figure \ref{fig:n_o}. Assuming ID60001 is compatible with the local N/O - O/H relation before dilution, it would require the dilution by a factor of $>5$ from the infall of pristine gas into a $12+\log({\rm O/H})\gtrsim 8.5$ galaxy, which is exceptionally high for the given $M_*$ at $z\sim4-5$ \citep[e.g.,][]{nakajima23,morishita24b}. As such, we find that scenario i) is unlikely to explain the enhanced N/O abundance in ID60001. 

For scenario iii), the depletion of Oxygen through SN winds would also result in the O/H abundance deficit similar to scenario i), hence is also disfavored. Notably, we do find outflow features represented by the broad components in [O{\sc iii}], H$\beta$, and H$\alpha$ emission lines (Figure \ref{fig:emlinefit}), as well as the spatially extended ionized gas (Figure \ref{fig:emsb}). With outflow components only detected in strongest emission lines due to the limited S/N of NIRSpec MSA data, we estimate the O/H abundance of the outflow component using the strong line method calibrated by \citet{sanders24}. Based on the $R3 = $[O{\sc iii}]$\lambda5007 / {\rm H\beta}$ flux ratio of $4.7\pm0.4$ for the outflow component of ID60001, we obtain two solutions of $12+\log({\rm O/H})\sim 7.4$ and $\sim8.4$, due to the bifurcating curve of the calibration. The latter solution is, again, considered exceptionally high for the given redshift, and thus scenario iii) is unlikely at face value. 
%Assuming the outflow of ID60001 has  $12+\log({\rm O/H})\sim8.4$, we can estimate the Oxygen depletion effect based on the ionized gas mass of the outflow ($M_{\rm gas,out}$) and the host galaxy ($M_{\rm gas,host}$). We estimate $M_{\rm gas,out}$, with the following equation based on H$\alpha$ luminosity of the outflow component ($L_{\rm H\alpha,{out}}$):
%\begin{equation}
    %M_{\rm gas,out} = \frac{1.36m_{\rm p}L_{\rm H\alpha,{out}}}{\gamma_{\rm H\alpha}n_{\rm e}},
%\end{equation}
%where $m_{\rm p}$, $n_{\rm e}$, and $\gamma_{\rm H\alpha}$ represent the atomic mass of Hydrogen, electron density of the outflow gas, and H$\alpha$ volume emissivity, respectively. We adopt $\gamma_{\rm H\alpha}=3.56\time10^{-25}~{\rm erg~cm^{-3}~s^{-1}}$ under the assumption of the case B recombination and electron temperature of $T_{\rm e}=10^4$~K. With $n_{\rm e}$ ranging from $100-1000~{\rm cm^{-3}}$, we obtain $\log(M_{\rm gas,out}/M_\odot)\sim 7.3-8.3$. For $M_{\rm gas,host}$, we adopt the average ionized gas to stellar mass ratio of $\sim 10$ for high-$z$ galaxies \citep{dg24}, resulting in $\log(M_{\rm gas,host}/M_\odot)=10.3$, a factor of $100-1000$ larger than $M_{\rm gas,out}$. From the O/H abundances and gas masses of the host galaxy and outflow gas, we estimate the Oxygen depletion effect of a $12+\log({\rm O/H})=8.4$ outflow is only by a factor of $< 0.15$, which is negligible.
% We thus conclude that scenario iii) is unlikely but cannot be completely ruled out. 
Deeper observations with higher S/N that capture outflow components in other emission lines are needed for a more robust estimation on the metallicity of the outflow component and hence determining whether or not scenario iii) can explain the high N/O abundance ratio in ID60001.

For scenario ii), it has been suggested that WR stars are able to expel Nitrogen into the ISM through stellar winds \citep[e.g.][]{cameron23,watanabe24} before they explode as core-collapse SNe (CCSNe) and quickly enrich the ISM with Oxygen. The time scale for elevated N/O abundance ratio is thus within the lifetime of WR phases (i.e. 3-4~Myr). However, \citet{watanabe24} further show that if WR stars in the mass range of $25-120~M_\odot$ with low metallicity and heavy Fe core directly collapse into black holes without undergoing the CCSNe phase, the time scale for elevated N/O abundance ratio can be extended to $\sim10$~Myr. To estimate the stellar age of ID60001, we compare the EW$_0$ of narrow H$\alpha$ emission line to the relevent model from \texttt{STARBURST 99} model, where the EW$_0({\rm H}\alpha)=800.0_{-31.7}^{+51.5}$~\AA\ corresponds to the stellar age of $\log({\rm age/[yr]})\leq6.8$~Myr. Such a time scale is  
consistent with the elevated N/O abundance ratio produced by direct collapse WR stars. 
%In fact, the model with $\sim80\%$ of WR stars undergoing direct collapse can explain the high N/O ratio of ID60001.
%%%%%%%%%%%%%%%%%%%%%%%%%%%%%%%%%%
%\begin{figure}[htb!]
%\begin{center}
%\includegraphics[scale=0.5]{wr.png}
%\end{center}
%\caption{N/O ratio as a function of stellar age for WR stars with CCSNe (black curve) and direct collapse (blue curves) from \citet{watanabe24}. The green dashed lines indicate a separate model from \citet{meynet06} with different fraction of direct collapse WR stars, where the horizontal extents illustrate the lifetime of 30-90~$M_\odot$ stars. ID60001 (red square) can be explained by the model with $\sim80\%$ WR stars undergoing direct collapse.
%}\label{fig:wr}
%\end{figure}
%%%%%%%%%%%%%%%%%%%%%%%%%%%%%%%%%%

We find several indirect evidences that support the existence of WR stars in ID60001. First, we detect a moderately broad (FWHM$=410$~km~s$^{-1}$) He{\sc ii}$\lambda$4686 emission line whose relative strength to H$\beta$ (Figure \ref{fig:bpt}) and EW$_0=6.7_{-1.0}^{+1.1}$ is consistent with typical galaxies that contain WR stars \citep[e.g.][]{guseva00,martin23}. We notice that other emission lines (i.e. [Fe{\sc iii}]$\lambda$4658) in the typical WR ``blue bump'' feature are not detected in our spectra. This is likely due to the combined effect of the intrinsically weak WR features in metal-poor systems \citep[e.g.][]{crowther06,shirazi12,morishita24a} and the limited S/N of our data. Second, we detect wind features that can be explained by WR stars in NIRSpec spectra around [O{\sc iii}]$\lambda\lambda$4959, 5007, H$\beta$, and H$\alpha$ emission lines (Figure \ref{fig:emlinefit}) with outflow velocities of $v_{\rm out}=250-340$~km~s$^{-1}$. In NIRCam photometry (Figure \ref{fig:galfit}), we also detect disturbed residuals {\it only} in the filters covering H$\beta$+[O{\sc iii}] and H$\alpha$ in our morphology analyses, indicating the potential existence of shock-ionized gas from stellar wind associated with WR stars.

\subsection{Effect of density on N/O measurement}
%%%%%%%%%%%%%%%%%%%%%%%%%%%%%%%%%%
\begin{figure*}[ht!]
\begin{center}
\includegraphics[scale=0.68]{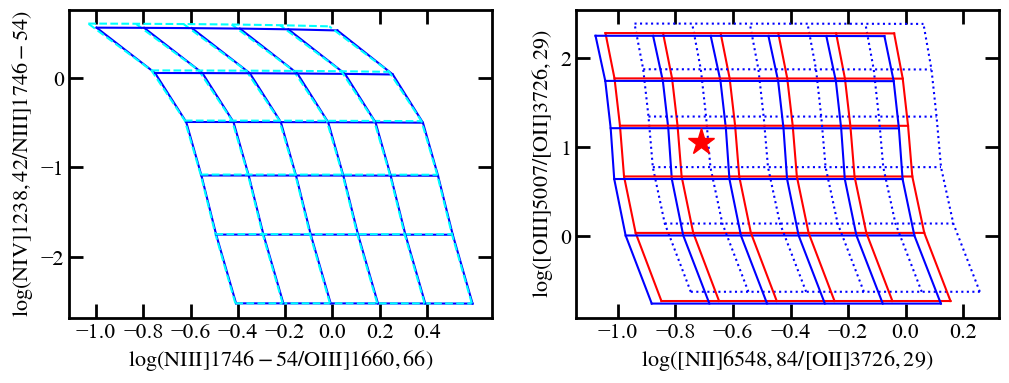}
\end{center}
\caption{Cloudy models for $n_{\rm H}=100$~cm$^{-3}$ (blue solid lines), 400~cm$^{-3}$ (red solid lines), 2000~cm$^{-3}$ ( blue dotted lines), and 20000~cm$^{-3}$ (cyan dashed lines). Each grid spans the N/O and ionization parameter ($U$) range of $-1.0\leq\log({\rm N/O})\leq 0.0$ and $-3.5\leq\log(U)\leq -1.0$, respectively. 
}\label{fig:cloudy}
\end{figure*}
%%%%%%%%%%%%%%%%%%%%%%%%%%%%%%%%%%

We investigate the robustness of our N/O measurement with Cloudy photoionization models \citep{cloudy23}. To model young star-forming galaxies, we use BPASS \citep{bpass18} binary stellar radiation assuming an instantaneous star formation history with the stellar age of 10~Myr, upper star-mass cut of 100~$M_\odot$, and the Salpeter IMF. We fix the stellar metallicity to nebular metallicity defined by O/H ratio derived in Section \ref{subsec:abundance}. We vary the N/O abundance ratio and ionization parameter ($U$) within the range of $-1\leq \log({\rm N/O})\leq 0$ and $-3.5\leq\log U\leq -1$. For hydrogen density ($n_{\rm H}$), we adopt different values of $n_{\rm H}=100, 400, 2000, 20000$~cm$^{-3}$ that correspond to the cases with low density, $n_{\rm e}$ value of ID60001, high density in singly-ionized Nitrogen regions, and high density in doubly/triply-ionized Nitrogen regions, respectively.

From the photoionization models, we predict the rest-frame UV and optical emission line ratios that are frequently adopted throughout N/O abundance derivations. 
%In the UV (optical) wavelength range, the N{\sc iii}]$\lambda\lambda$1746-54/O{\sc iii}]$\lambda\lambda$1660,66 ([N{\sc ii}]$\lambda$6548/[O{\sc ii}]$\lambda\lambda$3727,29) line ratio is often used to indicate the N/O abundance due to the similar ionization potential of N and O, while the N{\sc iv}]$\lambda\lambda$1238,42/N{\sc iii}]$\lambda\lambda$1746-54 ([O{\sc iii}]$\lambda$5007/[O{\sc ii}]$\lambda\lambda$3727,29) line ratio represents the ionization state. 
In the left panel of Figure \ref{fig:cloudy}, we demonstrate the predicted UV line ratios of N{\sc iii}]$\lambda\lambda$1746-54/O{\sc iii}]$\lambda\lambda$1660,66 versus N{\sc iv}]$\lambda\lambda$1238,42/N{\sc iii}]$\lambda\lambda$1746-54, where the lowest ($n_{\rm H}=100$~cm$^{-3}$) and highest ($n_{\rm H}=20000$~cm$^{-3}$) density grids overlap with each other. Such results indicate that the high ionization UV lines are relatively insensitive to the density variation, which has been mentioned in previous studies focusing on UV Nitrogen and Oxygen emission lines \citep[e.g.][]{isobe23c,ji24}. In contract, for the optical line ratios of [N{\sc ii}]/[O{\sc ii}] versus [O{\sc iii}]/[O{\sc ii}] (right panel of Figure \ref{fig:cloudy}), the high-density grid with $n_{\rm H}=2000$~cm$^{-3}$ exhibits a systematic shift for $\sim0.2$~dex in both positive vertical and horizontal axes compared with the low density grid ($n_{\rm H}=100$~cm$^{-3}$), suggesting that the gas density has a non-negligible effect on these optical emission line ratios. Thanks to the high S/N of NIRSpec MSA data, we are able to estimate the reliable $n_{\rm e}$ of ID60001 and avoid such systematics. For other galaxies whose N/O abundances are derived with rest-frame optical lines, it is necessary to obtain reliable $n_e$ measurements to avoid large systematics.

\section{Summary}\label{sec:summary}
In this paper, we present the ISM and host analysis on ID60001, a star-forming galaxy at $z=4.6927$, based on JWST NIRSpec MSA spectroscopy and NIRCam multiband imaging. From NIRSpec MSA spectra we have identified more than 30 emission lines, allowing us to measure the ISM properties and exclude the existence of AGN. Analyses on NIRCam imaging has yield to the derivation of host galaxy properties. Our main findings are summarized as follows:

i) We have derived the N/O and N/H abundance ratios from the rest-frame optical [N{\sc ii}]$\lambda$6584 emission line. Both abundance ratios are considerably elevated at the given gas-phase metallicity compared with local counterparts, indicating that ID60001 has an enriched Nitrogen abundance with respect to Oxygen and Hydrogen. We found evidence of outflow in a few bright emission lines. The ISM properties derived with and without the outflow component differ significantly (e.g., $\Delta A_V\sim1$). Caution is warranted for spectroscopic analysis based on low signal-to-noise ratio data or lower-resolution spectra. 

ii) We have conducted SED fitting to NIRCam multiband photometry 
%and NIRSpec MSA spectra simultaneously 
with \texttt{PROSPECTOR} and derived the physical properties of the host galaxy. The best-fit result has $\log(M_*/M_\odot) = 9.34_{-0.08}^{+0.10}$ and SFR = $13.49_{-2.32}^{+1.94}~M_\odot~{\rm yr}^{-1}$ during the most recent 10~Myr in lookback time. Multiband morphology analyses based on \texttt{Galfit} and spatial emission line maps suggest that ID60001 has an irregular morphology that can be best-described by two S\'ersic profiles, while the strong emission of H$\beta+$[O{\sc iii}] and H$\alpha$ are likely to have turbulent and spatially extended features. These results indicate that ID60001 features ionizing gas outflows originated from stellar winds.

iii) We have examined the possible explanations to the enriched Nitrogen abundance in ID60001, including dilution due to prinstine gas infall, Nitrogen enrichment from WR stars, and Oxygen depletion through SN winds. The most likely explanation is the existence of direct collapse WR stars that expel Nitrogen into the ISM without going through the CCSN phase which would enrich the ISM with Oxygen. The recent bursty SF history with a young stellar age, the detection of high-ionization, moderately broadened He{\sc ii} emission line from stellar ionizing sources, and the gas turbulence and outflow features originated from stellar wind all provide supporting evidence to the direct-collapse WR scenario for explaining the Nitrogen enrichment in ID60001.

iv) We demonstrated with Cloudy photoionization model that N/O abundance derived from optical emission lines are sensitive to the gas density. As such, the reliable estimation of electron density is crucial to the N/O measurement based on optical emission lines.

We emphasize that the conclusions of this paper is only based on the single object, while there may be multiple physical mechanisms that can explain Nitrogen enrichment in the other high-$z$ galaxies. Future observations that covers both UV and optical Nitrogen lines, as well as Carbon lines would provide key information to distinguishing the different Nitrogen enrichment mechanisms. 

%% Also note that the akcnowlodgment environment does not support long amounts of text. If you have a lot of people and institutions to acknowledge, do not use this command. Instead, create a new \section{Acknowledgments}.
\section*{Acknowledgements}
We thank Abdurro'uf and Zhaoran Liu for the discussion on the SED fitting analysis.
Some/all of the data presented in
this paper were obtained from the Mikulski Archive for Space
Telescopes (MAST) at the Space Telescope Science Institute. TM and YZ acknowledge the support provided by NASA through grant No. JWST-GO-3990 from the Space Telescope Science Institute, which is operated by AURA, Inc., under NASA contract NAS 5-26555. We acknowledge support for this work under NASA grant 80NSSC22K1294. 

\vspace{5mm}
\facilities{JWST(NIRCam and NIRSpec)}

\software{astropy \citep{astropy13,astropy18}, PROSPECTOR \citep{prospect}, 
          Cloudy \citep{cloudy23}, 
          SExtractor \citep{sex96}, numpy, PyNeb \citep{pyneb15} 
          }

%\appendix
%\section{Appendix information}

\bibliography{bib}{}
\bibliographystyle{aasjournal}
\end{document}